\documentstyle[12pt]{article}
\parskip=\medskipamount

\title{Eleven dimensional superstring with new supersymmetry and $D=10$
type IIA Green--Schwarz superstring}
\author{A.A. Deriglazov\thanks{deriglaz@phys.tsu.tomsk.su}}
\date{Department of Mathematical Physics,\\
Tomsk Polytechnical University, 634004 Tomsk, Russia}
\begin{document}
\maketitle
\large
\begin{abstract}
A covariant action for closed $D=11$ superstring with local
$\kappa$-symmetry and global supersymmetry transformations obeying
the algebra $\{Q_\alpha,Q_\beta\}=C\Gamma^{\mu\nu}P_\mu n_\nu$ is suggested.
Physical sector variables of the model and their dynamics exactly coincide
with those of the $D=10$ type IIA Green--Schwarz superstring. It is shown
that action of the $D=10$ type IIA Green--Schwarz superstring can be
considered as a partially gauge fixed version of the $D=11$ superstring
action.
\end{abstract}

\noindent
{\bf PAC codes:} 11.17.+y; 02.40.+m; 04.20.Jb\\
{\bf Keywords:} superstring, new higher dimensional superalgebras.

\section{Introduction}

While type IIA string and $p$-brane dualities (see [1, 2] and references
therein) indicate on a possibility of M-theory unification in (10,1)
dimensions [3, 4], theories which do not admit a direct M-theory unification
may arise from F-theory in (10,2) dimensions [5, 6]. Motivated by the
development of the F-theory, the authors of the recent works [7--13] have
suggested a number of various models in a space with signature $(D-2,2)$.
An interesting point is that they are based not on the super Poincar\'e
algebra but on some other one, with commutator of supersymmetry generators
of the type
\begin{equation}
\{Q_\alpha,Q_\beta\}\sim \Gamma^{\mu\nu}P_\mu n_\nu.
\end{equation}
In particular, for the case of superparticle (superstring) models the
algebra of such a type can be realized in a superspace as follows:
\begin{equation}
\delta\theta=\epsilon, \qquad \delta x^\mu=i\bar\epsilon
\Gamma^{\mu\nu}n_\nu\theta.
\end{equation}
To find interpretation for the vector $n^\mu$, it was suggested to consider
a system with two superparticles [7, 12, 13]. Then $P^\mu$ and $n^\mu$ may
be regarded as momenta for each member of the system.

In this letter another interpretation of the algebra (1) and the vector
$n^\mu$ will be presented for the case of $D=11$ space with standard
signature (10,1). Namely, we suggest a Poincar\'e invariant action
for $D=11$ superstring which is invariant under ``new supersymmetry''
[7--9] transformations (2), as well as under some additional bosonic
transformations, whose role is to provide on-shell closure of the
full algebra. The action presented includes a space-like vector $n^\mu$
as an auxilliary variable, which turns out to be gauged away. (On this
reason, it is not necessary to consider a pair of superstrings in our
construction.) Since the variable $n^\mu$ is treated on equal footing
with other ones, the symmetry transformations form a superalgebra in
the usual sence (without occurence of nonlinear in generator terms in
the right hand side of Eq. (1)), in contrast to Refs. 7, 12, 13.
Further, one possible gauge is $n^\mu=(0,\dots,0,1)$. In this gauge
Eq. (2) reduces (in our $\Gamma$-matrix notations [14]) to
\begin{equation}
\begin{array}{l}
\delta\theta_\alpha=\epsilon^\alpha, \qquad \delta\bar\theta_\alpha=
\bar\epsilon^\alpha,\\
\delta x^{\bar\mu}=-i\bar\epsilon_\alpha \tilde\Gamma^{\bar\mu\alpha
\beta}\bar\theta_\beta-i\epsilon^\alpha\Gamma^{\bar\mu}_{\alpha\beta}
\theta^\beta, \qquad \delta x^{11}=0,\end{array}
\end{equation}
where $\theta=(\bar\theta_\alpha,\theta^\alpha)$, $\mu=(\bar\mu,11)$,
$\bar\mu=0,1,\dots,9$, $\alpha=1,\dots,16$. Eq. (3) exactly coincides
with the standard $D=10,N=2$ supersymmetry transformations. Thus, in
our case one can treat the new supersymmetry (2) as a way to rewrite
the $D=10,N=2$ supersymmetry in the ``eleven dimensional notations''.

The work is organized as follows. In Sec. 2 we present and discuss
a model of a nondynamical space-like vector $n^\mu$, which
seems to be a necessary part of our construction suggested in the next
Section. In Sec. 3 covariant action for closed $D=11$ superstring is
suggested.
Its global symmetries are founded and prove to form an on-shell closed
algebra. Generalized local $\kappa$-symmetry is also presented. In Sec.
4 within the Hamiltonian framework it is shown that physical sector
variables and their dynamics coincide with those of the $D=10$ type IIA
Green--Schwarz (GS) superstring [15]. Thus, one gets the corresponding
supersymmetric spectrum on the quantum level. In Sec. 5 it is
demonstrated that the action for $D=10$ type IIA GS superstring can be
considered as a partially gauge fixed version of the $D=11$ superstring
action.

\section{Action for a nondynamical space-like vector}

As was mentioned in the Introduction, we need to get in our
disposal a nondynamical space-like vector field. So, as a preliminary
step of our construction, let us discuss the following $D=11$
Poincar\'e invariant action
\begin{equation}
S=-\int d^2\sigma \left[ n^\mu\varepsilon^{ab}\partial_a A^\mu_b+
\frac 1\phi (n^2+1)\right],
\end{equation}
which turns out to be a building block of the eleven dimensional
superstring action considered below. Here $n^\mu(\sigma^a)$ is
$D=11$ vector and $d=2$ scalar, $A^\mu_a(\sigma^b)$ is $D=11$ and $d=2$
vector, while $\phi(\sigma^a)$ is a scalar field. In Eq. (4) we have set
$\varepsilon^{ab}=-\varepsilon^{ba}$, $\varepsilon^{01}=-1$,
$\eta_{\mu\nu}=(+,-,\dots,-)$ and it is also supposed that
$\sigma^1\subset [0,\pi]$. From the equation of motion $\delta
S/\delta\phi=0$ it follows that $n^\mu$ is a space-like vector.

Local symmetries of the action are the $d=2$ reparametrizations\footnote{Note
that interaction with the $d=2$ metric $g^{ab}(\sigma^c)$ is not
necessary due to the presence of $\varepsilon^{ab}$ symbol and
the supposition that the variable $\phi$ transforms as a density
$\phi'(\sigma')={\rm det}(\partial\sigma'/\partial\sigma)\phi(\sigma)$
under the reparametrizations.} and the following transformations with
the parameters $\rho^\mu(\sigma^a)$, $\omega_a(\sigma^b)$
\begin{equation}
\begin{array}{l}
\delta A^\mu_a=\partial_a\rho^\mu+\omega_an^\mu;\\
\delta\phi=\displaystyle\frac 12 \phi^2\varepsilon^{ab}
\partial_a\omega_b.\end{array}
\end{equation}
These symmetries are reducible because their combination with
the parameters of a special form: $\omega_a=\partial_a\omega$,
$\rho^\mu=-\omega n^\mu$ is a trivial symmetry: $\delta_\omega
A^\mu_a=-\omega\partial_an^\mu$, $\delta_\omega\phi=0$ (note that
$\partial_an^\mu=0$ is one of the equations of motion). Thus, Eq.
(5) includes 12 essential parameters which correspond to the primary first
class constraints $p^\mu_0\approx0$, $\pi_\phi\approx0$ in the Hamilton
formalism (see below).

Let me demonstrate a nondynamical character of the model. For this aim the
Hamiltonian formalism seems to be the most appropriate, since second
class constraints must be taken into account. Momenta conjugate to the
variables $n^\mu$, $A^\mu_a$, $\phi$ are denoted by $p^\mu_n$,
$p^\mu_a$, $\pi_\phi$. All equations for determining the momenta turn
out to be the primary constraints
\begin{eqnarray}
&& \pi_\phi=0, \qquad p^\mu_0=0;\\
&& p^\mu_n=0, \qquad p^\mu_1-n^\mu=0.
\end{eqnarray}
The canonical Hamiltonian is
\begin{equation}
H=\int d\sigma^1\left[n^\mu\partial_1A^\mu_0+\frac 1\phi (n^2+1)
+\lambda_\phi\pi_\phi+\lambda^\mu_n p^\mu_n+\lambda^\mu_0
p^\mu_0+\lambda^\mu_1(p^\mu_1-n^\mu)\right],
\end{equation}
where $\lambda_*$ are the Lagrange multipliers corresponding to the
constraints. The preservation in time of the primary constraints
implies the secondary ones
\begin{equation}
\partial_1n^\mu=0, \qquad n^2+1=0,
\end{equation}
and equations for determining some of the Lagrange multipliers
\begin{equation}
\lambda^\mu_1=\partial_1 A^\mu_0+\frac 2\phi n^\mu, \qquad
\lambda^\mu_n=0.
\end{equation}
The tertiary constraints are absent.

Constraints (7) form a system of second class and can be omitted
after introducing the corresponding Dirac bracket (the Dirac
brackets for the remaining variables coincide with the Poisson ones).
After imposing the gauge fixing conditions $\phi=2$, $A^\mu_0=0$ for the
first class constraints (6), dynamics of the remaining variables is ruled
by the equations
\begin{equation}
\dot A^\mu_1=p^\mu_1, \qquad \dot p^\mu_1=0, \qquad
(p^\mu_1)^2=-1, \qquad \partial_1p^\mu_1=0.
\end{equation}
Then the gauge conditions $A^{11}_1=\tau$, $A^{\bar\mu}_1=0$ are
selfconsistent and lead to $p^{11}_1=1$, $p^{\bar\mu}_1=0$.
It should be stressed also that the equation $p^{11}_1=1$ is consistent
with the closed string boundary conditions only [14]. Hence, the model
(4) is selfconsistent being considered on the closed world sheet only.

Thus, we have demonstrated that one of the possible gauges to the
theory (4) is
\begin{equation}
A^{11}_1=\tau, \qquad p^{11}_1=n^{11}=1, \qquad \phi=2,
\end{equation}
with all other variables vanishing. Adding of this action to any model
is one of the ways to introduce (without change of the initial
dynamics) a space-like vector $n^\mu$, that may further be used
as appropriate. The action of such a kind was successfully used before
[16] in a different context.

\section{Action of $D=11$ superstring and its symmetries}

$D=11$ superstring action to be examined is of the form
\begin{eqnarray}
S=\int d^2\sigma\left\{\frac{-g^{ab}}{2\sqrt{-g}}\Pi_a^\mu\Pi_{b\mu}-
i\varepsilon^{ab}\Big(\partial_a x^\mu-\frac i2\bar\theta\Gamma^{\mu\nu}
n_\nu\partial_a\theta\Big)(\bar\theta\Gamma_\mu\partial_b\theta)-\right.\cr
\left.-\varepsilon^{ab}\xi_a(n_\mu\Pi^\mu_b)-n^\mu\varepsilon^{ab}
\partial_aA^\mu_b-\frac 1\phi (n^2+1)\right\},
\end{eqnarray}
where $\theta$ is a 32-component Majorana spinor of $SO(1,10)$, $\xi_a$
is a $d=2$ vector and it was denoted $\Pi^\mu_a\equiv \partial_a x^\mu
-i\bar\theta\Gamma^{\mu\nu}n_\nu \partial_a\theta$. The meaning of the
last two terms was explained in the previous section. The third term is
crucial for existence of local $\kappa$-symmetry and, at the same time,
it provides the split of the $x^{11}$ coordinate from the physical
sector (see below).

Let me describe global symmetries structure of the action (13). Bosonic
symmetries are the $D=11$ Poincar\'e transformations in the standard
realization and the following ones with antisymmetric parameter
$b^{\mu\nu}=-b^{\nu\mu}$:
\begin{equation}
\begin{array}{l}
\delta_bx^\mu={b^\mu}_\nu n^\nu,\\
\delta_bA^\mu_a=-{b^\mu}_\nu\displaystyle\left(\varepsilon_{ab}
\frac{g^{bc}}{\sqrt{-g}}{\Pi_c}^\nu-\xi_an^\mu+i(\bar\theta\Gamma^\nu
\partial_a\theta)\right).\end{array}
\end{equation}
There are also fermionic supersymmetry transformations being realized
as follows:
\begin{eqnarray}
&& \delta\theta=\epsilon,\cr
&& \delta x^\mu=i\bar\epsilon\Gamma^{\mu\nu}n_\nu\theta,\\
&& \delta A^\mu_a=i\varepsilon_{ab}\displaystyle\frac{g^{bc}}{\sqrt{-g}}
\Pi_{c\nu}(\bar\epsilon\Gamma^{\mu\nu}\theta)-\frac 56(\bar\epsilon
\Gamma^{\nu\mu}\theta)(\bar\theta\Gamma_\nu\partial_a\theta)+\cr
&&\qquad\qquad +\displaystyle\frac 16(\bar\epsilon\Gamma_\nu\theta)
(\bar\theta\Gamma^{\nu\mu}\partial_a\theta).\nonumber
\end{eqnarray}

One can prove that the complete algebra is on-shell closed up to the
equation of motion $\partial_an^\mu=0$ and trivial transformations
$\delta A^\mu_a=\partial_a\rho^\mu$ (see Eq. (5)) with field-dependent
parameter $\rho^\mu$, as it usually happens in component formulations of
supersymmetric models without auxilliary fields. The only nontrivial
commutator is\footnote{To elucidate relation between Eqs. (16) and (1)
let me point a simple analogy: algebra of the Lorentz generators
$M^{\mu\nu}=x^\mu p^\nu-x^\nu p^\mu$ can be written either as
$[M^{\mu\nu},M^{\rho\sigma}]=\eta^{\mu\rho}M^{\nu\sigma}+\dots$ or
$[M^{\mu\nu},M^{\rho\sigma}]=-\eta^{\mu\rho}p^\sigma x^\nu+\dots\,$.
The second case may be considered as corresponding to Eq. (1).}
\begin{equation}
[\delta_{\epsilon_1},\delta_{\epsilon_2}]=\delta_b, \qquad
b^{\mu\nu}=-2i(\bar\epsilon_1\Gamma^{\mu\nu}\epsilon_2).
\end{equation}
Let me note that one needs to use the $D=11$ Fierz identities to prove
Eq. (16) for $A^\mu_a$ variable
\begin{equation}
(C\Gamma^\mu)_{\alpha(\beta}(C\Gamma^{\mu\nu})_{\gamma\delta)}+
(C\Gamma^{\mu\nu})_{\alpha(\beta}(C\Gamma^\mu)_{\gamma\delta)}=0.
\end{equation}
The relation of Eq. (15) to the $D=10,N=2$ supersymmetry has been
described in the Introduction.

Local bosonic symmetries for the action (13) are the $d=2$
reparametrizations (with the standard transformation lows for all
variables except the variable $\phi$, which transforms as a density:
$\phi'(\sigma')={\rm det}(\partial\sigma'/
\partial\sigma)\phi(\sigma)$~), Weyl symmetry, and the transformations
with parameters $\rho^\mu(\sigma^a)$ and $\omega_a(\sigma^b)$ described
in the previous Section.

The action is also invariant under a pair of local fermionic
$\kappa$-symmetries. To find them, let me consider the following ansatz:
\begin{eqnarray}
&& \delta\theta=\pm\Pi_{d\mu}S^\pm\Gamma^\mu\kappa^{\mp d},\cr
&& \delta x^\mu=-\delta\bar\theta\Gamma^{\mu\nu}n_\nu\theta,\\
&& \delta g^{ab}=8i\sqrt{-g}P^{\pm ca}(\partial_c\bar\theta S^\mp
\kappa^{\mp b}),\nonumber
\end{eqnarray}
where it was denoted
\begin{equation}
S^\pm=\frac 12(1\pm n_\mu\Gamma^\mu), \quad
\kappa^{\mp d}\equiv P^{\mp dc}\kappa_c, \quad
P^{\mp dc}=\frac 12\left(\frac{g^{dc}}{\sqrt{-g}}\mp \varepsilon^{dc}
\right).
\end{equation}
Note that on-shell (where $n^2=-1$) the ${S^\pm}_\alpha{}^\beta$ --
operators form a pair of projectors in $\theta$-space. Let me recall
also that the $d=2$ projectors $P^\pm$ obey the following properties:
$P^{+ab}=P^{-ba}$, $P^{\mp ab}P^{\mp cd}=P^{\mp cb}P^{\mp ad}$.

After tedious calculations with the use of these properties and the
Fierz identities (17), a variation of the action (13) under the
transformations (18) can be presented in the form
\begin{equation}
\delta S=-\varepsilon^{ab}\partial_an_\nu G^\nu_b-\frac 1{\phi^2}
(n^2+1)H+\varepsilon^{ab}(n_\mu\Pi^\mu_b)F_a,
\end{equation}
where
\begin{eqnarray}
&&{G_b}^\nu\equiv -i\varepsilon_{bc}\displaystyle
\frac{g^{cd}}{\sqrt{-g}}(\delta\bar\theta\Gamma^{\mu\nu}\theta)
\Pi_{d\mu}+\frac 12 (\delta\bar\theta\Gamma^{\mu\nu}\theta)
(\bar\theta\Gamma_\mu\theta)-\cr
&&\qquad -\displaystyle\frac 12(\delta\bar\theta\Gamma_\mu\theta)
(\bar\theta\Gamma^{\mu\nu}\partial_b\theta)+i\xi_b
(\delta\bar\theta\Gamma^{\mu\nu}\theta)n_\nu,\cr
&& H\equiv -i\phi^2\displaystyle\frac{g^{ab}}{\sqrt{-g}}
(\partial_a\bar\theta\Gamma^\mu\tilde\kappa^\mp)\Pi_{b\mu},\\
&& F_a\equiv i\Big[\varepsilon_{ac}\displaystyle\frac{g^{cd}}
{\sqrt{-g}}(\partial_d\bar\theta\Gamma^\mu\tilde\kappa^\mp)n_\mu+
(\partial_a\bar\theta\tilde\kappa^\mp)\mp\cr
&&\qquad \mp 2\varepsilon_{ab}P^{\pm\,cd}(\partial_c\bar\theta\Gamma^\mu
\kappa^{\mp\,b})\Pi_{d\mu}\Big],\nonumber
\end{eqnarray}
and it was denoted $\tilde\kappa^\mp\equiv\Pi_{a\mu}\Gamma^\mu
\kappa^{\mp\,a}$. All the terms in Eq. (20) can evidently be cancelled
by corresponding variations of the auxiliary fields
\begin{equation}
\delta {A_b}^\nu={G_b}^\nu, \qquad \delta\phi=H, \qquad
\delta\xi_a=F_a.
\end{equation}

In the result, eleven dimensional superstring action (13) is invariant
under transformations from Eq. (18) supplemented by ones from Eq. (22).
Let me stress that all three last terms in the action turn out to be
essential for achieving this local $\kappa$-symmetry.

Since in Eq. (18) there appeared the double projectors ($S^\pm$ and
$\Pi_{a\mu}\Gamma^\mu$) acting on the $\theta$-space, the total number
of essential parameters is $8+8$. Their relation with the $D=10,N=2$ GS
superstring $\kappa$-symmetry will be described in the last Section.

\section{Analysis of dynamics}

The aim of this Section is to demonstrate that physical variables
of the theory (13) and their dynamics exactly coincide with those of
the $D=10$ type IIA GS superstring [15].

Following the standard Hamiltonian procedure one finds a pair of second
class constraints ${p_n}^\mu=0$, ${p_1}^\mu-n^\mu=0$ among primary
constraints of the theory. Then variables $(n^\mu,{p_n}^\mu)$ can be
omitted after introducing the associated Dirac bracket (see Sec. 2).
The Dirac brackets for the remaining variables coincide with the
Poisson ones, and the Hamiltonian with primary constraints then looks like
\begin{eqnarray}
& H=\displaystyle\int d\sigma^1\left\{-\frac N2(\hat p^2+\Pi_{1\mu}
\Pi_1^\mu)-N_1\hat p_\mu \Pi_1^\mu +p_{1\mu}\partial_1 A_0^\mu
-\xi_0(p_{1\mu}\partial_1x^\mu)+\right. \cr
& \left.+\displaystyle\frac 1\phi (p_1^2+1)+ \lambda_\phi\pi_\phi +
\lambda_{0\mu}p_0^\mu +\lambda^{ab}(\pi_g)_{ab}+\lambda_{\xi a}
{p_\xi}^a+L_\alpha{\lambda_\theta}^\alpha\right\},\label{ham}
\end{eqnarray}
where $p^\mu$, $p_0^\mu$, $p_1^\mu$, $p_{\xi a}$, $(\pi_g)_{ab}$
are momenta conjugate to the variables $x^\mu$, $A_0^\mu$, $A_1^\mu$,
$\xi_a$, $g_{ab}$, respectively; $\lambda_*$ are Lagrange multipliers
corresponding to the primary constraints. In Eq. (\ref{ham}) we also
denoted
\begin{equation}
\begin{array}{c}
N = \displaystyle\frac{\sqrt{-g}}{g^{00}}, \quad N_1=\frac{g^{01}}{g^{00}},
\quad \hat p^\mu=p^\mu-i\bar\theta\Gamma^\mu \partial_1\theta+
\xi_1p_1^\mu, \\
L_\alpha\equiv p_{\theta\alpha}-i\displaystyle\Big(p^\mu-\frac i2
\bar\theta\Gamma_\mu\partial_1\theta\Big)\bar\theta\Gamma^{\mu\nu}p_{1\nu}
-i\Big(\partial_1x^\mu-\frac i2 \bar\theta\Gamma^{\mu\nu}p_{1\nu}
\partial_1\theta\Big)\bar\theta\Gamma_\mu=0.\end{array}
\end{equation}

The full system of constraints can be presented in the form
$$
(\pi_g)_{ab}=0, \qquad \pi_\phi=0, \qquad p_{\xi a}=0, \qquad
p^\mu_0=0;
\eqno{(25.a)}$$
$$
\partial_1 p^\mu_1=0, \qquad (p^\mu_1)^2=-1;
\eqno{(25.b)}$$
$$
\hat p^\mu p_{1\mu}=0, \qquad \partial_1x^\mu p_{1\mu}=0;
\eqno{(25.c)}$$
$$
(\hat p^\mu\pm\Pi^\mu_1)^2=0, \qquad L_\alpha=0.
\eqno{(25.d)}$$
\addtocounter{equation}{1}
Besides, some of the Lagrange multipliers have been determined in the
process
\begin{equation}
\lambda^\mu_n=0, \qquad \lambda^\mu_{A1}=\partial_1 A^\mu_0+
\frac 2\phi p_1^\mu +Q^\mu,
\end{equation}
where
\begin{equation}
Q^\mu\equiv -N\xi_1\hat p^\mu+N_1\xi_1\Pi^\mu_1-\xi_0\partial_1x^\mu+
\frac 12[(\bar\theta\Gamma_\nu\partial_1\theta)\bar\theta\Gamma^{\mu\nu}
+(\bar\theta\Gamma^{\mu\nu}\partial_1\theta)\bar\theta\Gamma_\nu]
\lambda_\theta.
\end{equation}

To go further let me impose gauge fixing conditions to the first class
constraints (25.a). The choise consistent with the equations of motion
is
\begin{equation}
\begin{array}{c} g^{ab}=\eta^{ab}, \qquad \phi=2, \qquad \xi_a=0,\\
A^\mu_0=-\displaystyle\int\limits_0^\sigma d\sigma' Q^\mu(\sigma'),
\end{array}
\end{equation}
where $Q^\mu$ is given by Eq. (27). This choise for $A^\mu_0$
simplifies subsequent analysis of the $A^\mu_1,p^\mu_1$ sector. Namely,
dynamics of these variables is ruled now by the equations
\begin{equation}
\partial_0A^\mu_1=p_1^\mu, \qquad \partial_0 p^\mu_1=0,
\end{equation}
and by the first class constraints (25.b). The following gauge:
$A^{11}_1=\tau$, $A^{\bar\mu}_1=0$, can be
imposed, which breaks manifest $SO(1,10)$ covariance up to $SO(1,9)$
one. One gets also $p^{11}_1=1$, $p^{\bar\mu}_1=0$ and the
constraints (25.c) are reduced to $\hat p^{11}=0$, $\partial_1x^{11}=0$.
These are a pair of second class constraints which simply mean that the
variables $(x^{11},p^{11})$ can now be omitted.

In the result we stay with the situation of the $D=10$ GS superstring
(see Eq. (25.d)), and the subsequent analysis coincides with that well
known case [17, 18]. Namely, physical variables sector contains the
transverse components $x^i$, $i=1,\dots,8$, of the coordinate
$x^{\bar\mu}$, $\bar\mu=0,1,\dots,9$, and a pair of $SO(8)$ spinors
of opposite chirality $(\bar\theta_{\dot a},\theta_a)$,
$a,\dot a=1,\dots,8$.

They are related to the initial $\theta$-variable as follows:
\begin{equation}
\theta=\left(\begin{array}{c}\bar\theta_\alpha\\ \theta_\alpha
\end{array}\right), \quad \alpha=1,\dots,16, \quad \theta^\alpha=
\left(\begin{array}{c} S_a\\ \bar\theta_{\dot a}\end{array}\right),
\quad \bar\theta^\alpha=\left(\begin{array}{c} \theta_a\\ \bar S_{\dot
a}\end{array}\right).
\end{equation}
Dynamics of the physical variables
\begin{equation}
\begin{array}{l} \partial_0x^i=-p^i, \qquad
\partial_0p^i=-\partial_1\partial_1x^i,\\
(\partial_0+\partial_1)\theta_a=0, \qquad
(\partial_0-\partial_1)\bar\theta_{\dot a}=0,\end{array}
\end{equation}
as well as quantum states spectrum of the $D=11$ superstring (13)
exactly coincide with those of the $D=10$ type IIA GS superstring.

\section{Reduction to $D=10$ type IIA GS superstring action}

Type IIA string action can be considered as partially gauge fixed
version of the $D=11$ superstring (13), where $SO(1,10)$ invariance
is broken up to $SO(1,9)$. To demonstrate this, let me
substitute the gauge $n^\mu=(0,\dots,0,1)$, $\xi_a=0$ in Eq. (13) (then
equations of motion for $\xi_a$-variable mean that
$\partial_ax^{11}=0$). By using of $SO(1,9)$ decomposition for
$SO(1,10)$ objects [14]
\begin{eqnarray}
\begin{array}{l}
\Gamma^\mu=\{\Gamma^{\bar\mu},\Gamma^{11}\}=\left\{\left(\begin{array}{cc}
0 & \Gamma^{\bar\mu}\\ \tilde\Gamma^{\bar\mu} & 0\end{array}\right),
\left(\begin{array}{cc} 1 & 0\\ 0 & -1\end{array}\right)\right\},\\
\theta=(\bar\theta_\alpha,\theta^\alpha), \qquad
S^+=\left(\begin{array}{cc} 0 & 0\\ 0 & 1\end{array}\right), \qquad
S^-=\left(\begin{array}{cc} 1 & 0\\ 0 & 0\end{array}\right),\end{array}\\
\begin{array}{l}
\{\Gamma^{\bar\mu},\tilde\Gamma^{\bar\nu}\}=-2\eta^{\bar\mu\bar\nu},
\qquad \eta^{\bar\mu\bar\nu}=(+,-,\dots,-),\\
\alpha=1,\dots,16, \qquad \bar\mu=0,1,\dots,9,\\
\bar\theta\Gamma^{\mu\nu}n_\nu\psi=-\theta\Gamma^{\bar\mu}\psi-
\bar\theta\tilde\Gamma^{\bar\mu}\bar\psi,\\
\bar\theta\Gamma^\mu\psi=\{\bar\theta\tilde\Gamma^{\bar\mu}\bar\psi-
\theta\Gamma^{\bar\mu}\psi;\,-\theta\bar\psi-\bar\theta\psi\},
\end{array}\nonumber
\end{eqnarray}
the resulting expression can be rewritten (up to total derivative) in
the form
\begin{eqnarray}
\lefteqn{S =\int d^2\sigma\biggl\{\displaystyle\frac{-g^{ab}}{2\sqrt{-g}}
[\partial_ax^{\bar\mu}+i(\theta\Gamma^{\bar\mu}\partial_a\theta)+
i\bar\theta\tilde\Gamma^{\bar\mu}\partial_a\bar\theta]^2-}\cr
&& -i\varepsilon^{ab}\partial_ax^{\bar\mu}(\bar\theta\tilde\Gamma^{\bar\mu}
\partial_b\bar\theta-\theta\Gamma^{\bar\mu}\partial_b\theta)+
\varepsilon^{ab}(\theta\Gamma^{\bar\mu}\partial_a\theta)
(\bar\theta\tilde\Gamma^{\bar\mu}\partial_b\bar\theta)\biggr\},
\end{eqnarray}
which coincides with type IIA GS action [15]. In the similar fashion, global
supersymmetry transformations (15) reduces to the standard $N=2$
supersymmetry (3), as it was mentioned in the Introduction. At last, by
using of Eq. (32) the generalized $D=11$ $\kappa$-symmetry (18) reduces to
$D=10$ Siegel $\kappa$-symmetry of GS superstring action
\begin{eqnarray}
&& \delta\theta^\alpha=-P^{-cd}\Pi^{\bar\mu}{}_d \tilde\Gamma^{\bar\mu\,
\alpha\beta}\bar\kappa_{c\beta}, \qquad
\delta\bar\theta^\alpha=P^{+cd}\Pi^{\bar\mu}{}_d \Gamma^{\bar\mu}_
{\alpha\beta}{\kappa_c}^\beta,\cr
&& \delta x^\mu=i\theta^\alpha\Gamma^{\bar\mu}_{\alpha\beta}\delta
\theta^\beta+i\bar\theta_\alpha\tilde\Gamma^{\bar\mu\,\alpha\beta}
\delta\bar\theta^\beta,\\
&& \delta g^{ab}=8i\sqrt{-g}\{P^{-ca}(\partial_c\bar\theta\kappa^{+b})-
P^{+ca}(\partial_c\theta\bar\kappa^{-b})\}.\nonumber
\end{eqnarray}

In conclusion, in this letter $D=11$ Poincar\'e invariant superstring
action based on the new superslgebra (14)--(16) different from the super
Poincar\'e one was suggested. Physical sector variables, their dynamics
and states spectrum of the model coincide with one for the $D=10$ type
IIA GS superstring. In accordance with the results of Refs. 7 and 13 one
expects critical dimension of the theory is $D=11$. One may hope that
similar construction will works for lifting of the $D=10$ type IIB
string to corresponding (10,2) version (see also Ref. 13). It will be
interesting also to apply the scheme developed in this work for construction
of Lagrangean formulation for $(D-2,2)$ SYM equations of motion considered
in Refs. 9, 10. Note also that algebra of supersymmetry transformations is
closed on-shell only, and an intriguing problem is to find a formulation
with off-shell closed version of the superalgebra.

This work was supported by Joint DFG-RFBR project No 96-02-00180G.

\end{document}